\documentclass[a4paper,11pt,twoside]{article}
\usepackage[INRIA]{lipinria}

\usepackage[latin1]{inputenc}
\usepackage{a4wide}
\usepackage{url}
\usepackage{xspace}
\usepackage{subfigure}

\usepackage{figlatex}
\newcommand{\comm}{\textsf{$\delta$}} % comm for stage
\newcommand{\calc}{\textsf{w}}    % calc for  stage
\newcommand{\speed}{\textsf{s}}   % speed of proc
\newcommand{\bw}{\textsf{b}}      % bandwidth of link

\newcommand{\period}{T_{\textsf{period}}}
\newcommand{\latency}{T_{\textsf{latency}}}
\newcommand{\oppt}{T_{\textsf{opt}}}

\newcommand{\HETERO}{\textit{Fully Heterogeneous}\xspace}

\newcommand{\n}{n} %\textsf{n}} % number of stages
\newcommand{\p}{p} %\textsf{p}} % number of procs
\newcommand{\inn}{\textsf{in}}
\newcommand{\out}{\textsf{out}}
\newcommand{\first}{\textsf{first}}
\newcommand{\last}{\textsf{last}}

\newcommand{\link}{\textsf{link}} % number of procs
\renewcommand{\S}{\mathcal{S}} % stage
\usepackage{RR}
\begin{document}
\RRInumber{RR2008-2}
\RRNo{6410}
\RRItitle{Bi-criteria Pipeline Mappings \\for Parallel Image
  Processing}
\RRItitre{Mapping bi-cit\`ere de pipelines pour le traitement
  d'image en parall\`ele}
\RRIdate{January 2008}

\RRIauthor{Anne Benoit \and Harald Kosch \and Veronika Rehn-Sonigo \and Yves Robert}

\RRIthead{Bi-criteria Pipeline Mappings}
\RRIahead{A. Benoit \and H. Kosch \and V. Rehn-Sonigo \and Y. Robert}

\RRIabstract{
Mapping workflow applications onto parallel platforms is a challenging
problem, even for simple application patterns such as pipeline graphs.
Several antagonistic criteria should be optimized, such as throughput and
latency (or a combination). Typical applications include digital image
processing, where images are processed in steady-state mode.

In this paper, we study the mapping of a particular image
processing application, the JPEG encoding. Mapping pipelined JPEG encoding onto parallel
platforms is useful for instance for encoding Motion JPEG images. As the
bi-criteria mapping problem is NP-complete, we concentrate on the evaluation
and performance of polynomial heuristics.
}

\RRIresume{L'ordonnancement et l'allocation des workflows sur plates-formes
parall\`eles est un probl\`eme crucial, m\^eme pour des
applications simples comme des graphes en pipeline. Plusieurs crit\`eres contradictoires doivent \^etre
optimis\'es, tels que le d\'ebit et la latence (ou une combinaison
des deux). Des applications typiques incluent le traitement d'images
num\'eriques, o\`u les images sont trait\'ees en r\'egime permanent.

Dans ce rapport, nous \'etudions l'ordonnancement et l'allocation
d'une application de traitement d'image particuli\`ere, l'encodage
JPEG. L'allocation de l'encodage JPEG pipelin\'e sur des
plates-formes parall\`eles est par exemple utile pour l'encodage des
images Motion JPEG.  Comme le probl\`eme de l'allocation bi-crit\`ere
est NP-complet, nous nous concentrons sur l'analyse et \'evaluation
d'heuristiques polynomiales.
}

\RRIkeywords{pipeline, workflow application, multi-criteria
 optimization, JPEG encoding}
\RRImotscles{pipeline, application workflow, optimisation
 multi-crit\`ere, encodage JPEG}

\RRItheme{\THNum}
\RRIprojet{GRAAL}
\RRImaketitle

\section{Introduction}

This work considers the problem of mapping workflow
applications onto parallel platforms. This is a challenging
problem, even for simple application patterns. For homogeneous
architectures, several scheduling
and load-balancing techniques have been developed but the extension to heterogeneous clusters makes the
problem more difficult.

Structured programming
approaches rule out many of the problems which the
low-level parallel application developer is usually confronted to,
such as deadlocks or process starvation. We therefore focus on
pipeline applications, as they can easily be expressed as algorithmic
skeletons. More precisely, in this paper, we study the mapping of a particular pipeline
application: we
focus on the JPEG encoder (baseline process, basic mode). This image processing
application transforms numerical pictures from any format into a standardized
format called JPEG.
%This standard was developed almost 20 years ago to
%create a portable format for the compression of still images.
%We decided in favor of this application as in the theoretical point of
%view, it corresponds to a perfect pipeline workflow. %So we observe the
%behavior of our recently presented heuristics on a real life simulation.
%Meanwhile it has been extended to the JPEG 2000 standard to answer on missing
%requirements of today.
This standard was developed almost 20 years ago to create a portable
format for the compression of still images and new versions are
created until now (see http://www.jpeg.org/). JPEG (and later JPEG
2000) is used for encoding still images in Motion-JPEG (later
MJ2). These standards are commonly employed in IP-cams and are part of
many video applications in the world of game consoles. Motion-JPEG
(M-JPEG) has been adopted and further developed to several other
formats, e.g., AMV (alternatively known as MTV) which is a proprietary
video file format designed to be consumed on low-resource devices. 
The manner of encoding in M-JPEG and subsequent formats leads to a
flow of still image coding, hence pipeline mapping is appropriate.

We consider the different steps of the encoder as a
linear pipeline of stages, where each stage gets some input, has to
perform several computations and transfers the output to the next
stage. The corresponding mapping problem can be stated informally
as follows: which stage to assign to which processor? We require the
mapping to be interval-based, i.e., a processor is assigned an interval
of consecutive stages.
Two key optimization parameters emerge. On the one hand, we target
a high throughput, or short period, in order to be able to handle
as many images as possible per time unit. On
the other hand, we aim at a short response time, or latency, for the processing
of each image. These two criteria are antagonistic: intuitively,
we obtain a high throughput
with many processors to share the work, while we get a small latency by mapping
many stages to the same processor in order to avoid the cost of
inter-stage communications. 

%%  The
%% optimization problem is to determine the best mapping, over all possible
%% partitions into intervals, and over all processor assignments.
%% Based on the two parameters of optimization
%% that were evoced earlier, the objective can
%% be to minimize either the period, or the latency, or a combination: given a threshold
%% period, what is the minimum latency that can be achieved? and the counterpart:
%% given a threshold latency, what is the minimum period that can be
%% achieved?

%% commetnte, parce qu il n y a pas assez de place
The rest of the paper is organized as follows: Section~\ref{sec:JPEG}
 briefly describes JPEG coding principles. In Section~\ref{sec:framework}
 the theoretical and applicative framework is introduced, and
 Section~\ref{sec:LP} is dedicated to linear programming formulation of
 the bi-criteria mapping. In Section~\ref{sec:heuristics} we describe
 some polynomial heuristics, which we use for our experiments of
 Section~\ref{sec:simu}. We discuss related work in Section~\ref{sec:related}.
 Finally, we give some concluding remarks in Section~\ref{sec:conclusion}.

\section{Principles of JPEG encoding}
\label{sec:JPEG}

\begin{figure}
  \centering
   \includegraphics[width=1.0\textwidth]{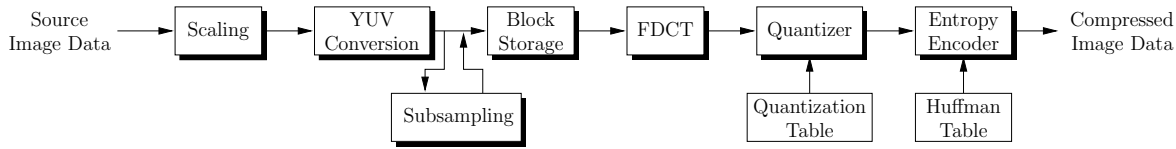}
  \caption{Steps of the JPEG encoding.}
  \label{fig:encoder}
\end{figure}

Here we briefly present the mode of operation of a JPEG
encoder (see~\cite{WallaceJPEG} for further details).
The encoder consists in %differentiates between 
seven pipeline stages, as shown in Figure~\ref{fig:encoder}.
%We suppose that the color space of
%the original picture is based on the RGB color
%model. %~\cite{RGBcolor}.
In the first stage, the image is scaled to
have a multiple of an 8x8 pixel matrix, and the standard even claims a multiple of 16x16.
In the next stage a color
space conversion is performed: the colors of the picture are
transformed from the RGB to the YUV-color model. %~\cite{RGBcolor}.
The sub-sampling
stage is an optional stage, which, depending on the sampling rate,
reduces the data volume: as the human eye can dissolve luminosity more
easily than color, the chrominance components are sampled more rarely
than the luminance components. Admittedly, this leads to a loss of data.
The last preparation step consists in the creation and storage of so-called
MCUs (Minimum Coded Units), which correspond to 8x8
pixel blocks in the picture.
The next stage is the core of the encoder. It performs a Fast
Discrete Cosine Transformation (FDCT) (eg.~\cite{FDCT}) on the 8x8
pixel blocks which are interpreted as a discrete signal of 64
values. After the transformation, every point in the matrix is
represented as a linear combination of the 64 points.
The quantizer reduces the image information to the important
parts. Depending on the quantization factor and quantization matrix,
irrelevant frequencies are reduced. Thereby quantization errors can
occur, that are remarkable as quantization noise or block generation
in the encoded image. The last stage is the entropy encoder, which
performs a modified Huffman coding: it combines the variable length
codes of Huffman coding with the coding of repetitive data in
run-length encoding.

\section{Framework}
\label{sec:framework}

\subsection{Applicative framework}
On the theoretical point of view, we consider a pipeline of $\n$
stages $\S_k$, $1 \leq k \leq \n$. %, as illustrated on
%Figure~\ref{fig.pipeline}.
Tasks are fed into the pipeline and processed from stage to
stage, until they exit the pipeline after the last stage.
The $k$-th stage $\S_k$ first receives an input from the previous
stage, of size $\comm_{k-1}$, then performs a number of $\calc_k$
computations, and finally outputs data of size $\comm_{k}$ to the
next stage. These three operations are performed sequentially. 
The first stage $\S_1$ receives an input of size $\comm_0$ from the
outside world, while the last stage $\S_{\n}$ returns the result, of
size $\comm_{\n}$, to the outside world, thus these particular stages
behave in the same way as the others.

On the practical point of view, we consider the applicative pipeline
of the JPEG encoder as presented in Figure~\ref{fig:encoder} and its
seven stages.

%% \begin{figure}[tbh]
%%   \begin{center}
%%     \includegraphics[width=0.7\textwidth]{fig/pipeline.fig}
%%   \end{center}
%%   \vspace{-0.4cm}
%%   \caption{The theoretical application pipeline.}
%%   \label{fig.pipeline}
%% \end{figure}

\subsection{Target platform}
We target a platform %(see Figure~\ref{fig.clique}),
with $\p$ processors
$P_u$, $1 \leq u \leq \p$, fully interconnected as a (virtual) clique. There is a
bidirectional link $\link_{u,v}: P_u \to P_v$ between any processor pair $P_u$ and $P_v$,
of bandwidth $\bw_{u,v}$.
The speed of processor $P_u$ is denoted as
$\speed_u$, and it takes $X/\speed_u$ time-units for $P_u$ to execute $X$ floating point
operations. We also enforce a linear cost model for communications, hence it takes
$X/\bw$ time-units to send (resp. receive) a message of size $X$ to (resp. from)
$P_v$. Communications contention is taken care of by enforcing the \emph{one-port}
model~\cite{BhatRagra03}.
%% In this model, a given processor can be
%% involved in a single communication at any time-step, either a send or a receive. However,
%% independent communications between distinct processor pairs can take place
%% simultaneously. The one-port model seems to fit the performance of some current MPI
%% implementations, which serialize asynchronous MPI sends as soon as message sizes exceed a
%% few megabytes~\cite{SaifPa2004}.

\subsection{Bi-criteria interval mapping problem} 
We seek to map
intervals of consecutive stages onto processors~\cite{subhlock-spaa96}.
Intuitively, assigning several consecutive
tasks to the same processor will increase their computational load, but may well
dramatically decrease communication requirements.
We search for a partition of $[1..\n]$ into $m \leq \p$
intervals $I_j = [d_j, e_j]$ such that $d_j \leq e_j$ for $1 \leq j \leq m$, $d_1 = 1$,
$d_{j+1}= e_j + 1$ for $1 \leq j \leq m-1$ and $e_m = \n$.

The optimization problem is to determine the best mapping, over all possible
partitions into intervals, and over all processor assignments. The objective can
be to minimize either the period, or the latency, or a combination: given a threshold
period, what is the minimum latency that can be achieved? and the counterpart:
given a threshold latency, what is the minimum period that can be
achieved?

The decision problem associated to this bi-criteria interval mapping
optimization problem is NP-hard, since the period minimization problem
is NP-hard for interval-based mappings (see~\cite{heteropar07}).

\section{Linear program formulation}
\label{sec:LP}

We present here an integer linear program to compute the optimal
interval-based bi-criteria mapping on \HETERO
platforms, respecting either a fixed latency or a fixed period. We assume $\n$ stages and $\p$ processors, plus two fictitious extra stages
$\S_0$ and $\S_{\n+1}$ respectively assigned to $P_{\inn}$ and $P_{\out}$. First we need
to define a few variables:

%\begin{itemize}
  \noindent For $k \in [0..\n+1]$ and $u \in [1..\p] \cup \{\inn, \out\}$,
  $x_{k,u}$ is a boolean variable equal to $1$ if stage $\S_k$
  is assigned to processor $P_u$; we let $x_{0,\inn} = x_{\n+1,\out} = 1$,
  and $x_{k,\inn} = x_{k,\out} = 0$ for $1 \leq k \leq \n$.

  \noindent For $k \in [0..\n]$, $u, v \in [1..\p] \cup \{\inn, \out\}$
  with $u \neq v$, $z_{k,u,v}$ is a boolean variable equal to $1$ if stage $\S_k$ is assigned to $P_u$
  and stage $S_{k+1}$ is assigned to $P_v$:
  hence $\link_{u,v}: P_u \to P_v$ is used for the communication
  between these two stages.
  If $k \neq 0$ then $z_{k,\inn,v}= 0$ for all $v \neq \inn$
  and if $k \neq \n$ then $z_{k,u,\out}= 0$ for all $u \neq \out$.

  \noindent For $k \in [0..\n]$ and  $u \in [1..\p] \cup \{\inn, \out\}$, $y_{k,u}$
  is a boolean variable equal to $1$ if stages $\S_k$
  and $S_{k+1}$ are both assigned to $P_u$; we let $y_{k,\inn} = y_{k,\out} = 0$
  for all $k$, and $y_{0,u} = y_{\n,u} = 0$ for all $u$.

  \noindent For $u \in [1..\p]$, $\first(u)$ is an integer variable
  which denotes the first stage assigned to $P_u$; similarly,
  $\last(u)$ denotes the last stage assigned to $P_u$. Thus
  $P_u$ is assigned the interval $[\first(u),\last(u)]$.
  Of course $1 \leq \first(u) \leq \last(u) \leq \n$.

  \noindent $\oppt$ is the variable to optimize, so depending on the
    objective function it corresponds either to the period or to the latency.
 %\end{itemize}

We list below the constraints that need to be enforced. For simplicity, we write $\sum_u$
instead of $\sum_{u \in [1..\p] \cup \{\inn, \out\}}$ when summing over all processors.
First there are constraints for processor and link usage:
%\begin{itemize}

\noindent Every stage is assigned a processor: $\forall k \in [0..\n+1], \qquad \sum_{u} x_{k,u} = 1$.

\noindent Every communication either is assigned a link or collapses
because both stages are assigned to the same processor:
  $$\forall k \in [0..\n], \qquad\sum_{u \neq v} z_{k,u,v}
  + \sum_{u} y_{k,u} = 1$$

\noindent If stage $\S_k$ is assigned to $P_u$ and stage $\S_{k+1}$ to $P_v$, then $\link_{u,v}: P_u \to P_v$
is used for this communication:
   $$\forall k \in [0..\n], \forall u,v \in [1..\p] \cup \{\inn,\out\}, u \neq v,
   \qquad x_{k,u} + x_{k+1,v} \leq 1 + z_{k,u,v}$$

\noindent If both stages $\S_k$ and $\S_{k+1}$ are assigned to $P_u$, then $y_{k,u} = 1$:
   $$\forall k \in [0..\n], \forall u \in [1..\p] \cup \{\inn,\out\},
   \qquad x_{k,u} + x_{k+1,u} \leq 1 + y_{k,u}$$

 \noindent If stage $\S_k$ is assigned to $P_u$, then necessarily
$\first_u \leq k \leq \last_u$. We write this constraint as:
  $$\forall k \in [1..\n], \forall u \in [1..\p], \qquad \first_u \leq k.x_{k,u} + \n.(1 - x_{k,u})$$
  $$\forall k \in [1..\n], \forall u \in [1..\p], \qquad \last_u \geq k.x_{k,u}$$

\noindent If stage $\S_k$ is assigned to $P_u$ and stage $\S_{k+1}$ is
assigned to $P_v\neq P_u$ ({\em i.e.,} $z_{k,u,v}=1$) then necessarily
$\last_u \leq k$ and $\first_v \geq k+1$ since we consider intervals.
We write this constraint as:
   $$\forall k \in [1..\n -1], \forall u,v \in [1..\p], u\neq v, \qquad \last_u \leq k.z_{k,u,v} + \n.(1 - z_{k,u,v})$$
  $$\forall k \in [1..\n -1], \forall u,v \in [1..\p], u\neq v, \qquad  \first_v \geq (k+1).z_{k,u,v}$$

\noindent The latency of schedule is bounded by $\latency$:

 and $t\in [1..p]\cup\{\inn, \out\}$. $$
  \sum_{u=1}^{p} \sum_{k=1}^{n}\left[
    \left(\sum_{t\neq u}
    \frac{\comm_{k-1}}{\bw_{t,u}}z_{k-1,t,u}\right) +
    \frac{\calc_k}{\speed_u} x_{k,u} \right] + \left(\sum_{u\in [1..p]
    \cup\{\inn\}} \frac{\comm_n}{\bw_{u,\out}} z_{n,u,\out}\right)
    \leq \latency $$
 and $t\in [1..p]\cup\{\inn, \out\}$.
 %% $$
%%   \sum_{u=1}^{p} \sum_{k=1}^{n}\left[
%%     \left(\sum_{t\neq u, t\in [1..p]\cup\{\inn, \out\}}
%%     \frac{\comm_{k-1}}{\bw_{t,u}}z_{k-1,t,u}\right) +
%%     \frac{\calc_k}{\speed_u}\right] + \left(\sum_{u\in [1..p]
%%     \cup\{\inn\}} \frac{\comm_n}{\bw_{u,\out}} z_{n,u,\out}\right)
%%     \leq \latency $$

\noindent There remains to express the period of each processor and to
constrain it by $\period$:

$\forall u \in [1..\p],$\\
$$\sum_{k=1}^{n} \left\{
\left( \sum_{t \neq u} \frac{\comm_{k-1}}{\bw_{t,u}} z_{k-1,t,u} \right) +
\frac{\calc_{k}}{\speed_u} x_{k,u} + \left( \sum_{v \neq u} \frac{\comm_{k}}{\bw_{u,v}}
z_{k,u,v} \right) \right\} \leq \period$$ %\cycletime_u$$
%\end{itemize}

Finally, the objective function is either to minimize the period
$\period$ respecting the fixed latency $\latency$ or to minimize the
latency $\latency$ with a fixed period $\period$. So in the first case
we fix $\latency$ and set $\oppt = \period$. In the second case
$\period$ is fixed a priori and  $\oppt = \latency$.
With this mechanism the objective function reduces to minimizing
$\oppt$ in both cases.

\section{Overview of the heuristics}
\label{sec:heuristics}

The problem of bi-criteria interval mapping of workflow applications is NP-hard~\cite{heteropar07},
so in this section we briefly describe polynomial heuristics to solve it. See~\cite{heteropar07} for a more
complete description or refer to the Web at:\\
\centerline{\url{http://graal.ens-lyon.fr/~vsonigo/code/multicriteria/}}

In the following, we denote by $\n$ the number of
stages, and by $\p$ the number of processors.
We distinguish two sets of heuristics. The heuristics of the first set aim
to minimize the latency respecting an a priori fixed
period. The heuristics of the second set minimize
the counterpart: the latency is fixed a priori and we try to
achieve a minimum period while respecting the latency constraint.

\subsection{Minimizing Latency for a Fixed Period}

All the following heuristics sort processors by non-increasing speed, and
start by assigning all the stages to the first (fastest)
processor in the list. This processor becomes \emph{used}.

\paragraph{\bf H1-Sp-mono-P: Splitting mono-criterion --}% This heuristic sorts the
%processors by decreasing speed, and starts by assigning all the stages to the first
%processor in the list. This processor becomes used. Then, at
At each step, we select the
used processor~$j$ with the largest period and we try to split its stage interval,
giving some stages to the next fastest processor $j'$ in the list (not yet used). This
can be done by splitting the interval at any place, and either placing the first part of
the interval on $j$ and the remainder on $j'$, or the other way round. The solution which
minimizes $max(period(j),period(j'))$ is chosen if it is better than the original
solution. Splitting is performed as long as we have not reached
the fixed period or until we cannot improve the period anymore.

\paragraph{\bf H2-Sp-bi-P: Splitting bi-criteria --} This
heuristic uses a binary search over the latency. For this purpose at
each iteration we fix an authorized
increase of the optimal latency (which is obtained by mapping all stages on
the fastest processor), and
we test if we get a feasible solution via splitting. The splitting
mechanism itself is quite similar to {\bf H1-Sp-mono-P} except
that we choose the solution that minimizes $max_{i
\in\{j,j'\}}(\frac{\Delta latency}{\Delta period(j)})$ within the
authorized latency increase to decide where to split.
While we get a feasible solution, we reduce the authorized latency
increase for the next iteration of the binary search, thereby aiming at minimizing the mapping global latency.

\paragraph{\bf H3-3-Sp-mono-P: 3-splitting mono-criterion --} At each step
we select the used processor~$j$ with the largest period and we split its interval into three parts.
For this purpose we try to map two parts of the interval on the next pair of fastest processors in the list, $j'$ and $j''$, and to keep the third part on processor $j$. Testing all possible permutations and all possible positions where to cut, we choose the solution that minimizes $max(period(j),period(j'),period(j''))$.

\paragraph{\bf H4-3-Sp-bi-P: 3-splitting bi-criteria --} In this heuristic
the choice of where to split is more elaborated: it depends not only
of the period improvement, but also of the latency increase.
Using the same splitting mechanism as in {\bf H3-3-Sp-mono-P}, we select
the solution that minimizes  $max_{i\in\{j,j',j''\}}(\frac{\Delta
latency}{\Delta period(i)})$.
Here $\Delta latency$ denotes the difference between the global
latency of the solution before the split and after the split. In the
same manner $\Delta period(i)$ defines the difference between the
period before the split (achieved by processor $j$) and the new period
of processor $i$.

\subsection{Minimizing Period for a Fixed Latency}

 As
in the heuristics described above, first of all we sort processors
according to their speed and map all stages on the fastest processor.

\paragraph{\bf H5-Sp-mono-L: Splitting mono-criterion --} This heuristic uses
the same method as  {\bf H1-Sp-mono-P} with a different break
condition. Here splitting is performed as long as we do not exceed the
fixed latency, still choosing the solution that minimizes
$max(period(j),period(j'))$.

\paragraph{\bf H6-Sp-bi-L: Splitting bi-criteria --} This variant of the
splitting heuristic works similarly to {\bf H5-Sp-mono-L}, but at each
step it chooses the solution which minimizes
$max_{i\in\{j,j'\}}(\frac{\Delta latency}{\Delta period(i)})$ while
the fixed latency is not exceeded.

\paragraph{Remark}
In the context of M-JPEG coding, minimizing the latency for a fixed period
corresponds to a fixed coding rate, and we want to minimize the response
time. The counterpart (minimizing the period respecting a fixed
latency $L$) corresponds to the question: if I accept to wait $L$
time units for a given image, which coding 
rate can I achieve?
We evaluate the behavior of the heuristics with respect to these questions
in Section~\ref{sec:TH}.

\section{Experiments and simulations}
\label{sec:simu}

In the following experiments, we study the mapping of the JPEG application onto
clusters of workstations. % under different aspects.

\begin{figure}
  \centering

  \includegraphics[width=0.8\textwidth]{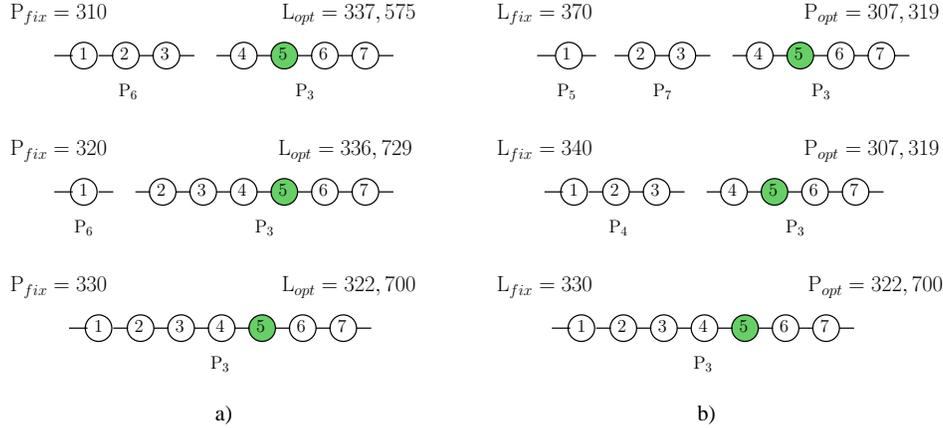}

  \caption{LP solutions strongly depend on
    fixed initial parameters.}
  \label{fig:pfix}
\end{figure}

\subsection{Influence of fixed parameters}

\begin{figure}
   \centering
   \subfigure[Fixed P.]{
     \includegraphics[angle=270,width=0.37\textwidth]{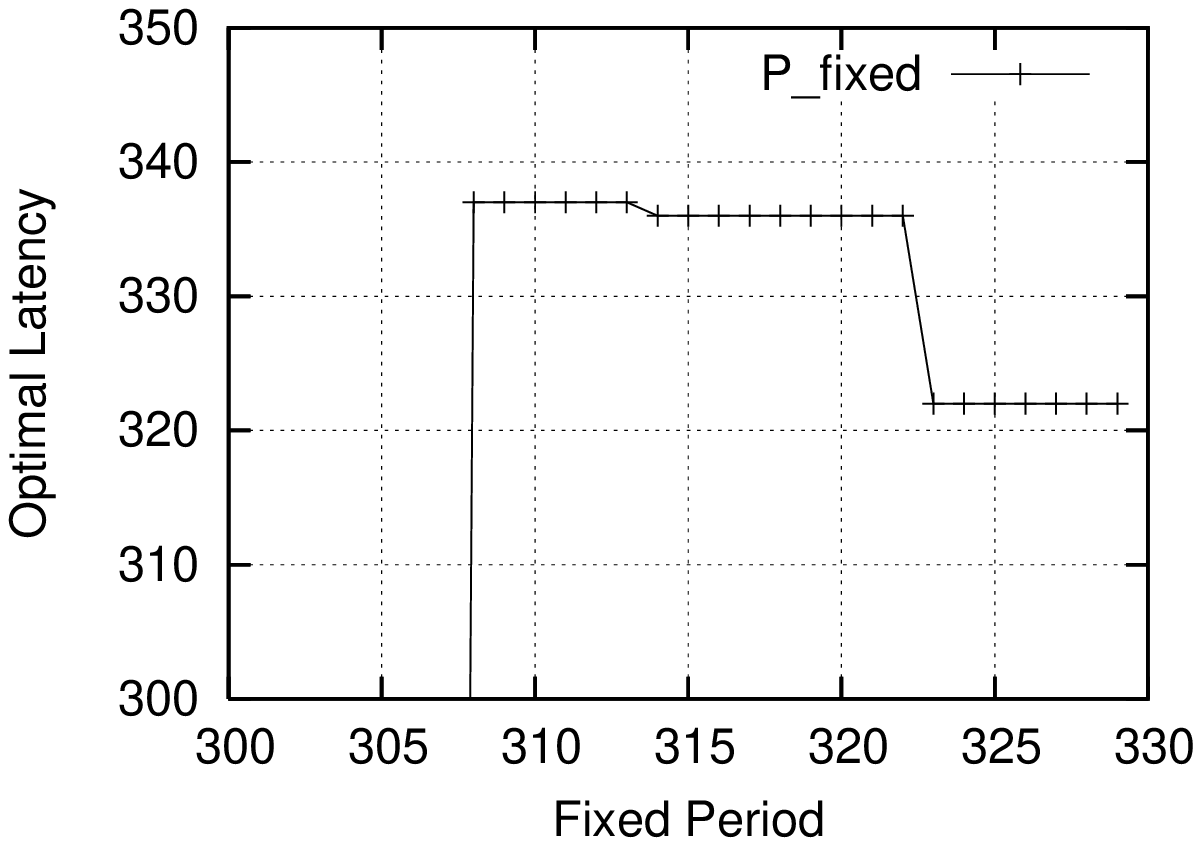}
     \label{fig:fixedP}
   }$\quad$
   \subfigure[Fixed L.]{
     \includegraphics[angle=270,width=0.37\textwidth]{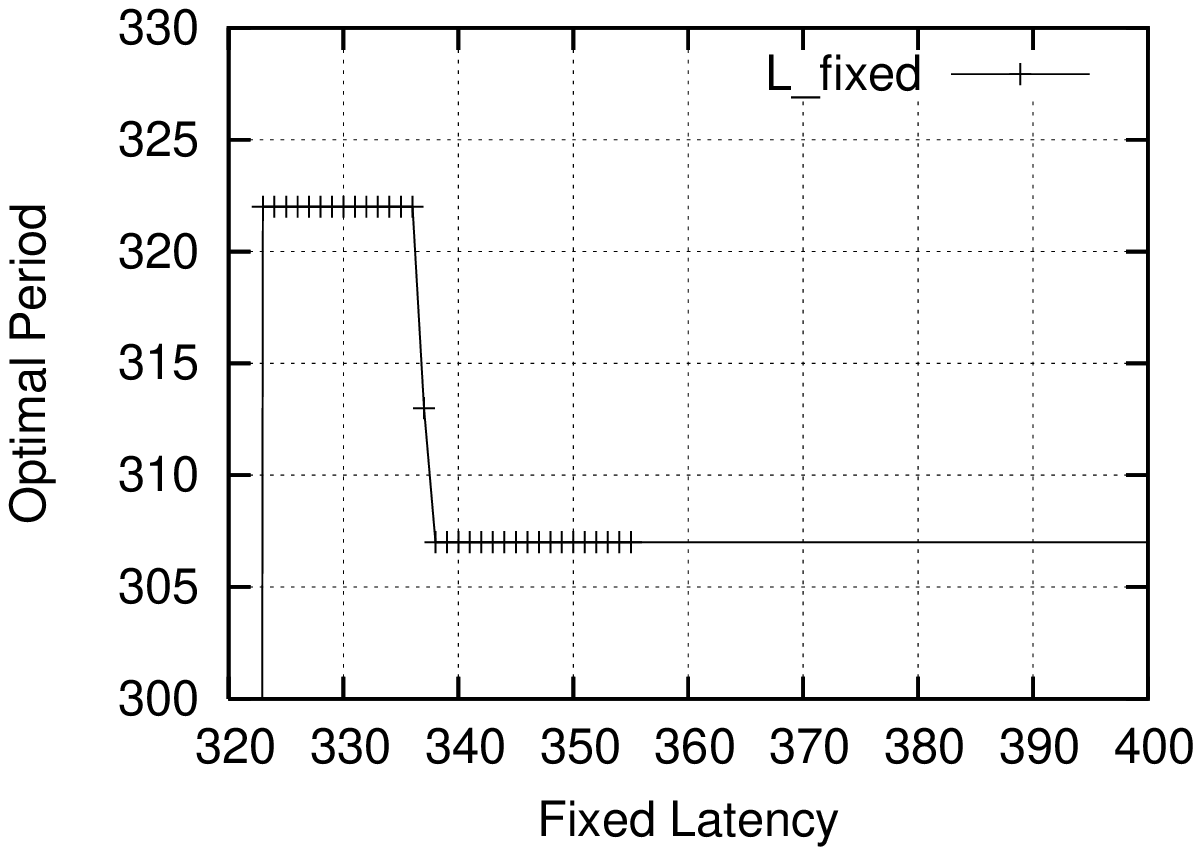}
     \label{fig:fixedL}
   }
   \caption{Bucket behavior of LP solutions.}
\label{fig:bucket}
\end{figure}

In this first test series, we examine the influence of fixed
parameters on the solution of the linear program. As shown in
Figure~\ref{fig:pfix}, the division into intervals is highly dependant
of the chosen fixed value. %The dark colored stage corresponds to the
%DCT stage, which takes the largest part of the computations.
The
optimal solution to minimize the latency (without any supplemental
constraints) obviously consists in mapping  the whole application pipeline
onto the fastest processor. As expected, if the period
fixed in the linear program is not smaller than the latter optimal mono-criterion latency,
this solution is
chosen. Decreasing the value for the fixed period imposes to split the stages among several
processors, until no more solution can be found.
Figure~\ref{fig:pfix}(a) shows the division into intervals for a fixed
period. A fixed period of $\period = 330$ is sufficiently high for
the whole pipeline to be mapped onto the fastest processor, whereas smaller periods lead
to splitting into intervals. We would like
to mention, that for a period fixed to $300$, there exists no solution anymore.
The counterpart - fixed latency - can be found in
Figure~\ref{fig:pfix}(b). Note that the
first two solutions find the same period, but for a different latency.
The first solution has a high value for latency, which allows
more splits, hence larger communication costs.
Comparing the last lines of Figures~\ref{fig:pfix}(a) and~(b), we state
that both solutions are the same, and we have $\period = \latency$.
Finally, expanding the range of the fixed values, a sort of bucket
behavior becomes apparent: Increasing the fixed parameter has in a
first time no influence, the LP still finds the same solution until
the increase crosses an unknown bound and the LP can find a better
solution. This phenomenon is shown in Figure~\ref{fig:bucket}.

% courbe avec nb od splits
% courbe avec les temps de L ou P trouves.
\subsection{Assessing heuristic performance}
\label{sec:TH}
\begin{figure}
   \centering
   \subfigure[Fixed P = 310.]{
     \includegraphics[angle=270,width=0.4\textwidth]{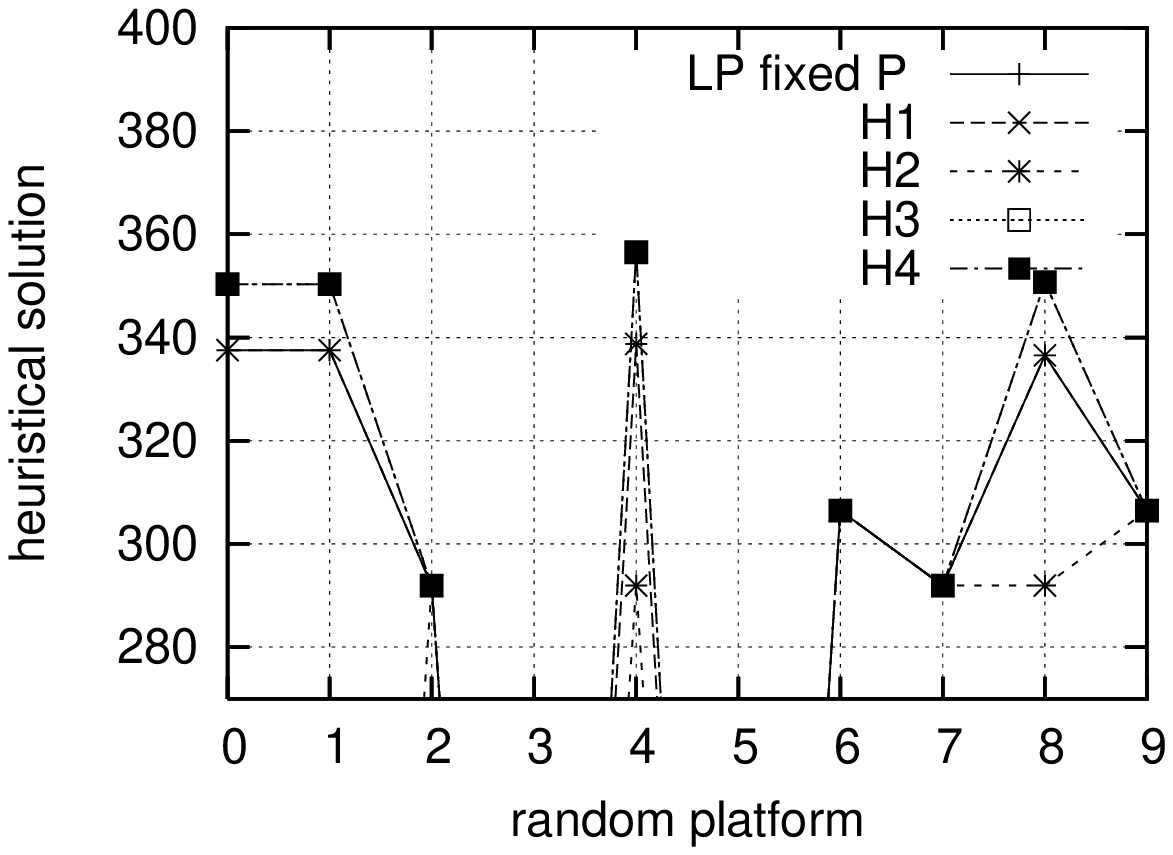}
     \label{fig:TH_P}
   }$\quad$
   \subfigure[Fixed L = 370.]{
     \includegraphics[angle=270,width=0.4\textwidth]{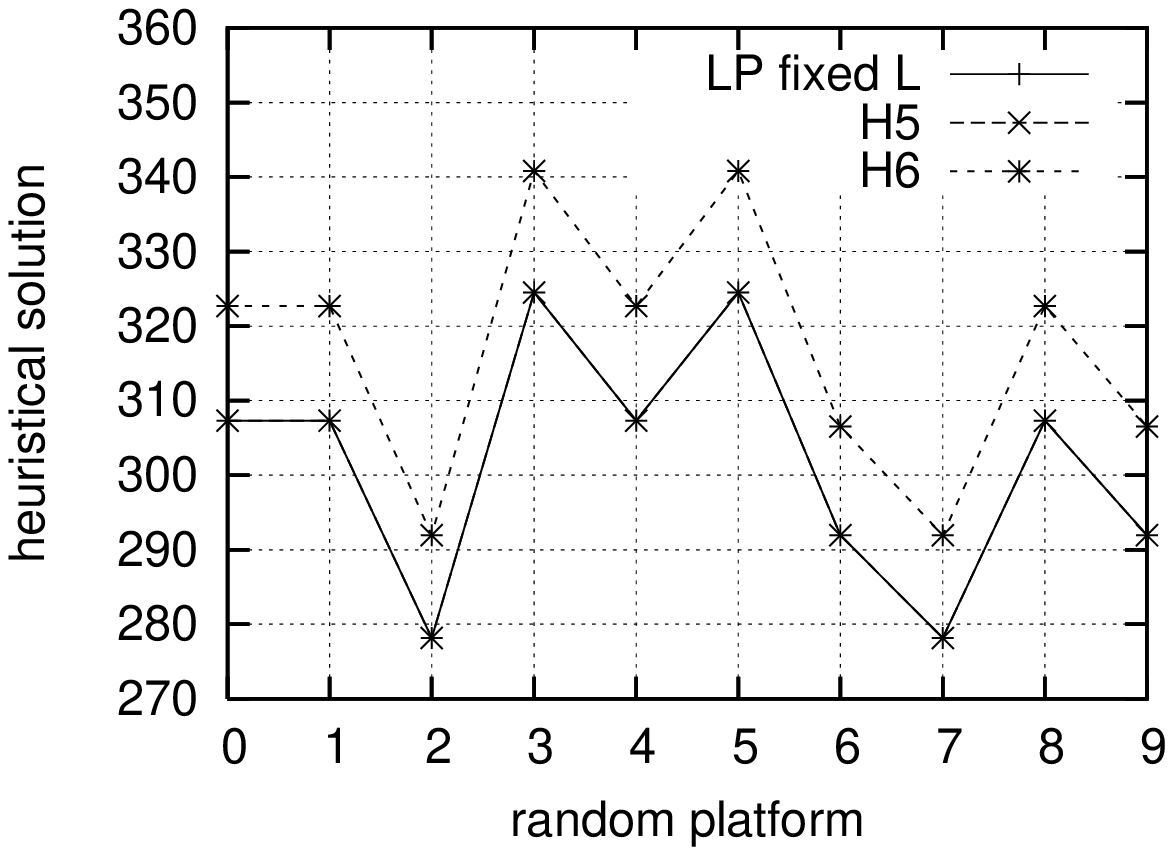}
     \label{fig:TH_L}
   }
   \caption{Behavior of the heuristics (comparing to LP solution).}
\label{fig:TH}
\end{figure}

The comparison of the solution returned by the LP program,
in terms of optimal latency respecting a fixed period (or the converse)
with the heuristics is shown in Figure~\ref{fig:TH}. The implementation is fed
with the parameters of the JPEG encoding pipeline and computes the
mapping on 10 randomly created platforms with 10 processors. On
platforms 3 and 5, no valid solution can be found for the fixed
period. There are two important
points to mention. First, the solutions found by H2 often are not
valid, since they do not
respect the fixed period, but they have the best ratio
latency/period. Figure~\ref{fig:h2} plots some more details: H2 achieves
good latency results, but the fixed period of P=310 is often
violated. This is a consequence of the fact that the fixed period value is very
close to the feasible period. When the tolerance for the period is
bigger, this heuristic succeeds to find low-latency solutions.
Second, all solutions, LP and heuristics, always keep the stages 4 to
7 together (see~Figure~\ref{fig:pfix} for an example). As stage 5 (DCT) is the most
costly in terms of computation, the interval containing these stages
is responsible for the period of the whole application.

Finally, in the comparative study H1 always finds the optimal period for
a fixed latency and we therefore recommend this heuristic for period
optimization. In the case of latency minimization for a fixed period, then H5
is to use, as it always finds the LP solution in the experiments.
This is a striking result, especially given the fact that the LP
integer program may require a  long time to compute the solution (up
to 11389 seconds in our experiments), while the heuristics always
complete in less than a second, and find the corresponding optimal
solution. 

\subsection{MPI simulations on a cluster}

This last experiment performs a JPEG encoding simulation. All
simulations are made on a cluster of 
homogeneous Optiplex GX 745 machines with an Intel Core 2
Duo 6300 of 1,83Ghz. Heterogeneity is enforced by increasing and
decreasing the number of operations a processor has to execute. The
same holds for bandwidth capacities.
We call this experiment simulation, as we do not parallelize a real
JPEG encoder, but we use a parallel pipeline application
which has the same parameters for communication and computation as the JPEG
encoder. An mpich implementation of MPI is used for communications.

\begin{figure}
   \centering

\subfigure[Simulative latency.]{
     \includegraphics[angle=270,width=0.4\textwidth]{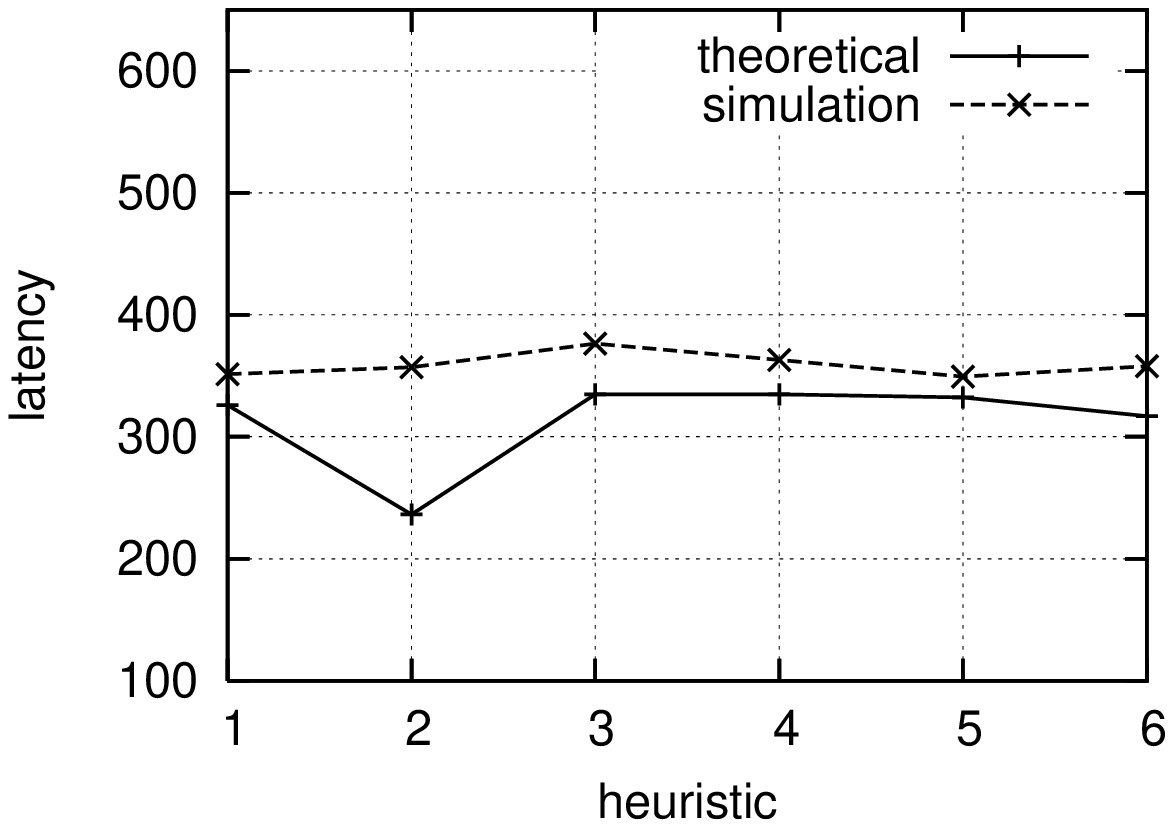}
     \label{fig:simu_L}
   }$\quad$
   \subfigure[H2 versus LP.]{
     \includegraphics[angle=270,width=0.4\textwidth]{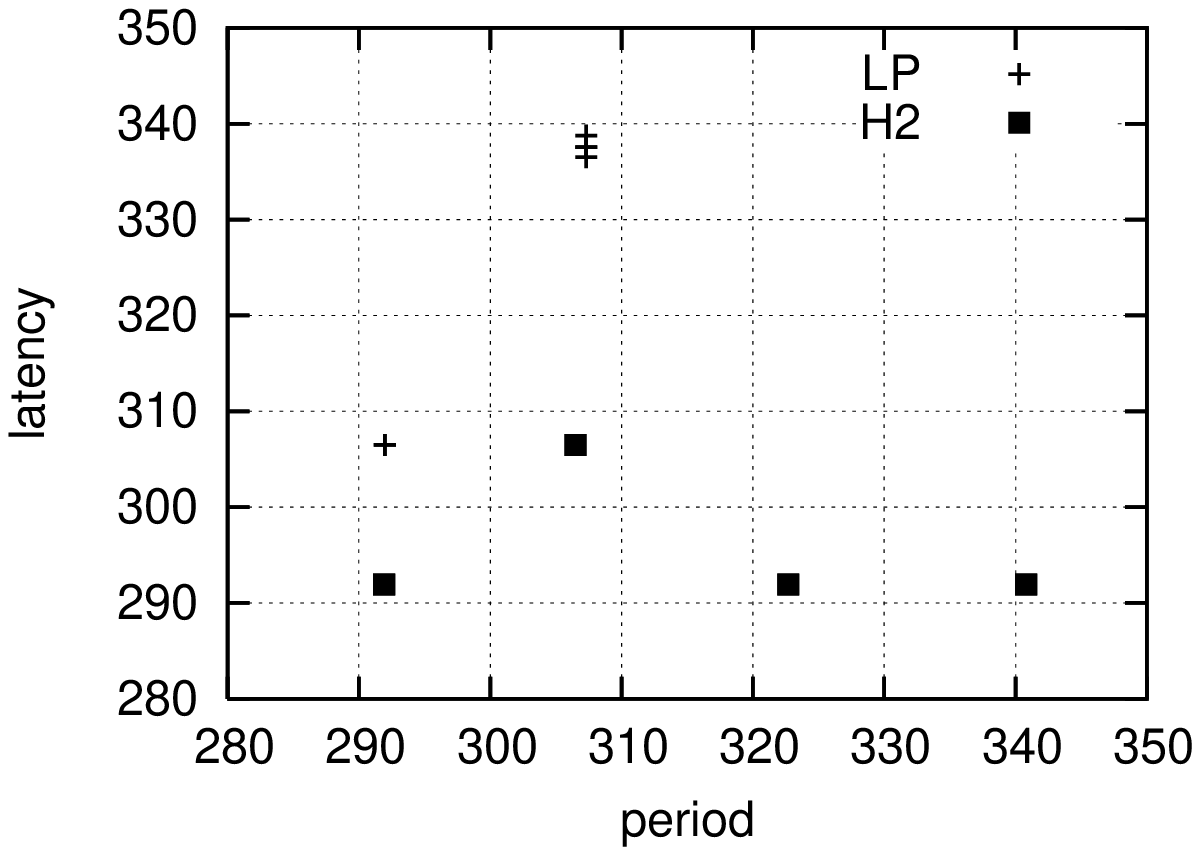}
     \label{fig:h2}
   }
   \caption{MPI simulation results.}
\label{fig:simu}
\end{figure}

In this experiment the same random platforms with 10 processors and
fixed parameters as in
the theoretical experiments are used. We measured the latency of the
simulation, even for the
heuristics of fixed latency, and computed the average over all random
platforms. Figure~\ref{fig:simu_L} compares the average of the
theoretical results of the heuristics to the average simulative
performance. The simulative
behavior nicely mirrors the theoretical behavior, with the exception of H2
(see Figure~\ref{fig:h2}). Here
once again, some solutions of this heuristic are not valid, as they do
not respect the fixed period. %Abstracting away from the latter, we conclude that our
%heuristics lead to a better solution than the naive approach.
%%YR which is represented by?

\section{Related work}
\label{sec:related}

The blockwise independent processing of the JPEG encoder allows to apply
simple data parallelism for efficient parallelization. Many papers have addressed
this fine-grain parallelization opportunity~\cite{Falkem95,ShenCook}. In addition, parallelization
of almost all stages, from color space conversion, over DCT to the Huffman
encoding has been addressed~\cite{Agostini,kuroda}. Recently, with respect to the JPEG2000 codec,
efficient parallelization of wavelet coding has been introduced \cite{JPEG2000}. All these
works target the best speed-up with respect to different architectures and
possible varying load situations. Optimizing the period and the latency is an
important issue when encoding a pipeline of multiple images, as for
instance for Motion JPEG (M-JPEG). To meet these issues, one has to solve in
addition to the above mentioned work a bi-criteria optimization problem, i.e.,
optimize the latency, as well as the period.
The application of coarse grain
parallelism seems to be a promising solution. We propose to use an
interval-based mapping strategy allowing multiple stages to be mapped to one
processor which allows meeting the most flexible the domain constraints (even
for very large pictures).
Several pipelined versions of the JPEG encoding have been considered. They rely mainly on pixel or block-wise parallelization~\cite{Ferretti,Papadonikolakis07}. For instance, Ferretti et al. [6] uses
three pipelines to carry out concurrently the encoding on independent pixels
extracted from the serial stream of incoming data. The pixel and block-based approach is however
useful for small pictures only.
Recently, Sheel et al.~\cite{Sheel07} consider a pipeline architecture where each stage presents a step in the JPEG encoding. The targeted architecture consists of Xtensa LX processors which run subprograms of the JPEG encoder program. Each program accepts data via the queues of the processor, performs the necessary computation, and finally pushes it to the output queue into the next stage of the pipeline. The basic assumptions are similar to our work, however no optimization problem is considered and only runtime (latency) measurements are available. The schedule is static and set according to basic assumptions about the image processing, e.g., that the DCT is the most complex operation in runtime.

\section{Conclusion}
\label{sec:conclusion}

In this paper, we have studied the bi-criteria (minimizing latency and period)
mapping of pipeline workflow applications, from both a theoretical and practical point of view.
On the theoretical side, we have presented an integer linear programming
formulation for this NP-hard problem. 
On the practical side, we have studied in depth the interval mapping
of the JPEG encoding pipeline on a 
cluster of workstations. Owing to the LP solution, we were able to
characterize a bucket behavior in the optimal solution, depending on
the initial parameters. Furthermore, we have compared  the behavior of some
polynomial heuristics to the LP solution and we
were able to recommended two heuristics with almost optimal behavior for parallel
JPEG encoding.
Finally, we evaluated the heuristics running
a parallel pipeline application with the same parameters as a JPEG encoder.
%and confirmed the superior behavior of the heuristics.
The heuristics were designed for general pipeline applications, and
some of them were aiming at applications with a large number of stages
(3-splitting), thus a priori not very efficient on the JPEG
encoder. Still, some of 
these heuristics reach the optimal solution in our experiments, which
is a striking result. %Future work include the design of new ad-hoc
%heuristics for this particular application, which we expect to be even

A natural extension of this work would be to consider further image processing
applications with more pipeline stages or a slightly more complicated pipeline
architecture. Naturally, our work extends to JPEG 2000 encoding which
offers among others wavelet coding and more complex multiple-component
image encoding~\cite{Christopoulos02}. 
Another extension is for the MPEG coding family which uses lagged
feedback: the coding of some types of frames depends on other
frames. Differentiating the types of coding algorithms, a pipeline
architecture seems again to be a promising solution architecture.

%\bibliographystyle{abbrv}

%\bibliography{biblio}

\begin{thebibliography}{10}

\bibitem{Agostini}
L.~V. Agostini, I.~S. Silva, and S.~Bampi.
\newblock {Parallel color space converters for JPEG image compression}.
\newblock {\em Microelectronics Reliability}, 44(4):697--703, 2004.

\bibitem{heteropar07}
A.~Benoit, V.~Rehn-Sonigo, and Y.~Robert.
\newblock {Multi-criteria Scheduling of Pipeline Workflows}.
\newblock In {\em {{HeteroPar'07}, Algorithms, Models and Tools for Parallel
  Computing on Heterogeneous Networks (in conjunction with Cluster 2007)}}.
  IEEE Computer Society Press, 2007.

\bibitem{BhatRagra03}
P.~Bhat, C.~Raghavendra, and V.~Prasanna.
\newblock Efficient collective communication in distributed heterogeneous
  systems.
\newblock {\em Journal of Parallel and Distributed Computing}, 63:251--263,
  2003.

\bibitem{Christopoulos02}
C.~Christopoulos, A.~Skodras, and T.~Ebrahimi.
\newblock {The JPEG2000 still image coding system: an overview}.
\newblock {\em IEEE Transactions on Consumer Electronics}, 46(4):1103--1127,
  2000.

\bibitem{Falkem95}
J.~Falkemeier and G.~Joubert.
\newblock Parallel image compression with jpeg for multimedisa applications.
\newblock In {\em High Performance Computing: Technologies, Methods and
  Applications}, Advances in Parallel Computing, pages 379--394. North Holland,
  1995.

\bibitem{Ferretti}
M.~Ferretti and M.~Boffadossi.
\newblock {A Parallel Pipelined Implementation of LOCO-I for JPEG-LS}.
\newblock In {\em 17th International Conference on Pattern Recognition
  (ICPR'04)}, volume~1, pages 769--772, 2004.

\bibitem{kuroda}
T.~Kumaki, M.~Ishizaki, T.~Koide, H.~J. Mattausch, Y.~Kuroda, H.~Noda,
  K.~Dosaka, K.~Arimoto, and K.~Saito.
\newblock {Acceleration of DCT Processing with Massive-Parallel Memory-Embedded
  SIMD Matrix Processor}.
\newblock {\em IEICE Transactions on Information and Systems - LETTER- Image
  Processing and Video Processing}, E90-D(8):1312--1315, 2007.

\bibitem{JPEG2000}
P.~Meerwald, R.~Norcen, and A.~Uhl.
\newblock {Parallel JPEG2000 Image Coding on Multiprocessors}.
\newblock In {\em {{IPDPS'02}, International Parallel and Distributed
  Processing Symposium}}. IEEE Computer Society Press, 2002.

\bibitem{Papadonikolakis07}
M.~Papadonikolakis, V.~Pantazis, and A.~P. Kakarountas.
\newblock {Efficient high-performance ASIC implementation of JPEG-LS encoder}.
\newblock In {\em Proceedings of the Conference on Design, Automation and Test
  in Europe {(DATE2007)}}, volume IEEE Communications Society Press, 2007.

\bibitem{Sheel07}
S.~L. Shee, A.~Erdos, and S.~Parameswaran.
\newblock {Architectural Exploration of Heterogeneous Multiprocessor Systems
  for JPEG}.
\newblock {\em International Journal of Parallel Programming}, 35, 2007.

\bibitem{ShenCook}
K.~Shen, G.~Cook, L.~Jamieson, and E.~Delp.
\newblock An overview of parallel processing approaches to image and video
  compression.
\newblock In {\em Image and Video Compression}, volume Proc. SPIE 2186, pages
  197--208, 1994.

\bibitem{subhlock-spaa96}
J.~Subhlok and G.~Vondran.
\newblock Optimal latency-throughput tradeoffs for data parallel pipelines.
\newblock In {\em {ACM} Symposium on Parallel Algorithms and Architectures
  {SPAA'96}}, pages 62--71. ACM Press, 1996.

\bibitem{WallaceJPEG}
G.~K. Wallace.
\newblock The {JPEG} still picture compression standard.
\newblock {\em Commun. ACM}, 34(4):30--44, 1991.

\bibitem{FDCT}
C.~Wen-Hsiung, C.~Smith, and S.~Fralick.
\newblock {A Fast Computational Algorithm for the Discrete Cosine Tranfsorm}.
\newblock {\em IEEE Transactions on Communications}, 25(9):1004--1009, 1977.

\end{thebibliography}

\end{document}